# Design Theory to improve health evidence retrieval

**Alvet Miranda and Shah Jahan Miah**

Newcastle Business School, The University of Newcastle, NSW, Australia.

Emails: contact@alvet.com.au; shah.miah@newcastle.edu.au

## ABSTRACT

**Objective:** Our study objective is to design a feasible technology solution for health organizations to remove barriers to evidence-based clinical information retrieval, and improve Evidence-Based Practice.

**Methods:** Literature from 2010 to 2020 was reviewed to define problems in evidence-based clinical information retrieval with recommendations from literature used to define solution objectives. Design Science Research is used to complete three projects in a research stream using cloud services such as Web-Scale Discovery, Content Management System, Federated Access, Global Knowledgebase, and Document Delivery. Design thinking, systems thinking, and user-oriented theory of information need are adopted to construct a design theory.

**Results:** The research stream produced three novel and innovative artefacts: a contextual model, a unified architecture, and a context aware unified architecture which we evaluate as part of academic reviews, scholarly publications, and conference proceedings in various research stream stages. A fourth artefact or design theory is presented to generalize results as mature knowledge.

**Conclusion:** Design theory provides practical knowledge for health sciences libraries to source existing cloud services or components and integrate them as a holistic solution at their health organizations. Design theory also contributes to academic knowledge by addressing pain points and 'call to action' identified during literature review.

# INTRODUCTION

Eight pain points or problems to accessing online clinical information were identified in a commentary titled "Why are they not accessing it? User barriers to clinical information access" (Laera et al., 2021). These pain points (Figure 1) comprise time, access, awareness, financial limitations, paywalls, integration, resource scope, and resource platforms. Each was identified by participants at a summit including publishers and health sciences librarians who manage their organization's online evidence resources.

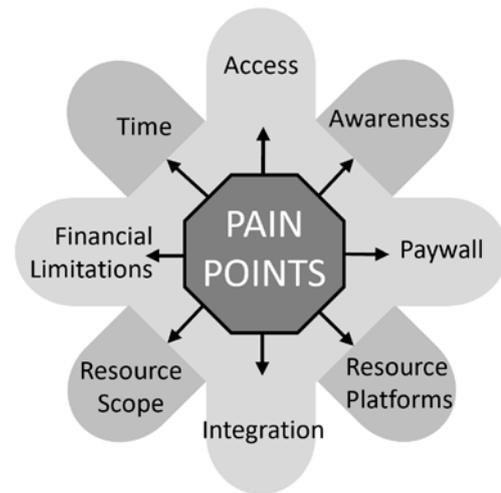

*Figure 1: Pain points to clinical information access.*

A 'call to action' was made to industry members and health sciences librarians to contribute mitigations to remove barriers to online information retrieval.

This identification of pain points to accessing evidence-based clinical information is consistent with academic literature over 10 years (2010-2020) that discusses the need to improve Evidence-Based-Practice (EBP) in healthcare organizations. EBP comprises patient values, clinical expertise, and research evidence, with access to clinical information being essential for clinicians to deliver patient care, conduct research, and to fill in knowledge gaps (Miranda & Miah, 2019).

**Purpose and Scope**

Our objective is to improve EBP in health organizations by removing barriers to clinical information retrieval. We target the research evidence component of EBP and use pain points to define problems. A survey of 439 nurses and physicians from public and private hospitals reported clinician's patient care improved when online clinical information (Lialiou et al., 2016). While use of online evidence-based systems are associated with Point-of-Care and Clinical Decision Support Systems, the survey results highlighted a different use case that involves writing scientific manuscripts and filling knowledge gaps. The former associated with clinical expertise and patient values components of EBP, and the latter supports research evidence.

Healthcare as a traditional industry is generally resistant to change and innovative practices with issues around information security, quality of care and privacy drawing worldwide attention (Maria & Shah, 2019). We aim to propose a design theory or generalized knowledge to facilitate implementation of systems by health sciences librarians through the sourcing and integration of existing cloud services in a way that addresses these pain points in an iterative and incremental manner for effective change management.

**Research Background**

Review of literature (2010-2020) regarding barriers to clinical informational retrieval and 'call to action' to improve research evidence identified issues and theoretical solutions to solve this 'wicked' (*sensu* Rittel and Webber) problem (Rittel & Webber, 1984). While attempts have been made to solve pain points in fragments (Miranda & Miah, 2021), no evidence of a holistic solution that addresses all pain points is known. A research design that enables health organizations to reconfigure existing systems to address pain points is required, which of course raises several questions: 1) do underlying theories exist that could be adopted to remove the barriers to clinical information; 2) what systems are used by health organizations to facilitate clinical information retrieval; and 3), could these systems be (feasibly) reconfigured by health sciences libraries in their respective organizations to remove barriers?

We took an iterative and incremental approach to solving these problems and designed three novel and innovative artefacts that were combined to produce a holistic solution. We designed and evaluated artefacts in a research stream consisting of three projects (A, B and C). We based evaluations on the FEDS framework (Venable et al., 2016) that offers four different strategies which dictate the scope of formative and summative evaluations either as artificial or naturalistic depending on an artefact's environment.

Project A focused on answering the first research question, to identify and adopt the theory of information need for information retrieval (Cole, 2011) discussed in the Methods section. The designed artefact was a user-oriented information retrieval contextual model using WSD to address the time, resource scope, and resource platforms pain points (Miranda & Miah, 2019). The model used filters or facets available in a WSD service, which were pre-set to apply context. Context allowed pre-setting a group context which could be done by librarians. Clinicians unfamiliar with the system could select or adopt a group context to get start and then define their own context settings as their system familiarity matured.

Usefulness of the context model was evaluated by interviewing clinicians and conducting content analysis of results at a confirmatory focus group in line with ethics approval for university level research. Project A results were compiled as a research article and published in a peer-reviewed open-access medical journal (Miranda & Miah, 2019).

Project B addressed the second question and designed a unified library services architecture to integrate cloud services in health organizations to solve the access, paywall, integration, and awareness pain points. All library systems at a national medical institute were identified and classified as one of resource provider or resource facilitator. The resulting architecture used a CMS as a front-end and integrated WSD, Global Knowledgebase, Link Resolver and Document Delivery (inter-library loan) via Application Programming Interface (API) for a unified user experience. The architecture was evaluated as useful by the library manager of the medical institution and a national conference review panel. The results were published as a book chapter after presentation at an annual national medical libraries conference in 2019 on site in New Delhi, India (Miranda, 2019).

Project C combined the two previous artefacts as a context aware unified architecture offering a holistic solution (Miranda, 2020) to answer the third question. The context aware unified architecture was first demonstrated to and then evaluated by an academic panel. We then adopted a multiple case studies method for further evaluation by conducting in-depth interviews (Hevner et al., 2004) with library managers at three different health organizations.

Case study research is well suited for information systems implementations when the context is important. A three-cluster framework for case study strategies was used to select the literal replication multi case design (Shakir, 2002) with three or four cases considered adequate for snowball (referrals by people interested in a case) or opportunistic (follow leads from field work) strategies.

We targeted health organizations as the natural environment for the designed architecture. Managers directly overseeing library services were identified as sources for data through in-depth interviews - one of the best means of informing case studies (Jafari Sadeghi et al., 2019).

The selection process was initiated by inviting participants (health sciences librarians) at the national medical conference during presentation of Project B to contribute to future research. Three organizations were selected through leads from the conference (opportunistic selection). One of the selected library managers had attended the conference presentation and another two leads were selected by referral from conference attendees (snowball selection).

On-location meetings were arranged by email including an overview of research and participation scope along with information on ethics approvals for consent to record the interviews for subsequent transcription purposes. Prior to recording any interview on the day, the research scope and ethics were explained again to proceed with the library manager's consent. An overview of the artefact architecture and a walkthrough of the architecture's instantiation were provided, inviting discussion for their specific case. Following the interview, recordings were transcribed, and scripts were imported into NVIVO for thematic analysis. Interviews were semi-structured with an agenda item to demonstrate the artefact and discussion during and after demonstration. This provided us with an opportunity to explore (Asher et al., 2013) and improve the artefact.

Results and outcomes from Project C were presented to librarians and online content providers at an international conference virtually in 2020 as plans for onsite presentation in London, UK were disrupted due to the COVID-19 pandemic. An open access copy of presentation slides and video recording were made available online after the conference (Miranda, 2020).

In addition, results were prepared as a research article and published in a peer-reviewed journal (Miranda & Miah, 2021). Findings of the context aware unified architecture were presented following the outline of Gregor and Hevner to communicate and present Design Science Research (DSR) for maximum impact (Gregor & Hevner, 2013). DSR contributions for publication range from specific, to limited to more abstract knowledge covering the spectrum of instantiations, constructs, methods, models, and design theories. According to Gregor's taxonomy the research output is an architecture that formalizes knowledge or theory for design and action on how to do something (Gregor, 2006).

Baskerville et. al. (Baskerville et al., 2018) distinguishes knowledge contributions between science and technology with the goal of science being to grow descriptive knowledge, and technology to grow prescriptive knowledge of designed artefacts. We lean towards technology producing a prescriptive design theory for knowledge contribution in publication. While each of the three artefacts was evaluated as useful in its own right, the implementation was based on an operational architecture or technological rules for health sciences libraries. We present well-developed implementation principles as mature and generalized knowledge in this paper.

## METHODS

The role of DSR is to identify problems, problem context, the audience and then iteratively and incrementally design an artifact for that audience to use as a solution. Significant rigor is applied during artefact design, and its instantiations are evaluated against requirements and results communicated to the scholarly body of knowledge or industry professionals (Johannesson & Perjons, 2014). DSR has proven to an effective research paradigm providing a new way of thinking in terms of making information systems design research relevant to practitioners (Genemo et al., 2016).

Artefacts from DSR can take many forms at varying levels: instantiations (software or process) are considered level 1, where an artefact is implemented; level 2 artefacts comprise constructs, methods, models and design principles produced as operational principles or architectures; and level 3 artefacts are more well-developed theories of embedded phenomena and take the form of design theories (Gregor & Hevner, 2013) which is the focus of this paper.

Construction of design theories requires instantiations to present a working artefact, either as a prototype or implementation with more complete, abstract and mature knowledge derived from more specific, less mature and limited knowledge forms (Gregor & Jones, 2007). Our research progressed in a research stream (Baskerville et al., 2018), comprising of three DSR projects that produced publication opportunities due to their novel and innovative artefact output. Each project informed the next, following a trajectory from specific instantiations of artefacts to artefacts as a model, and then artefacts as information system architectures. Results of these three artefacts in the research stream inform our design theory in line with DSR which in addition to being a methodology to develop artefacts to solve problems also enables researchers to learn from solutions (Miah et al., 2019a).

Each project used the Design Science Research Methodology (DSRM) (Peffers et al., 2008) to design artefacts with methods such as systems thinking (Kasser & Zhao, 2016) and design thinking (Corral & Fronza, 2018) being added as the projects became progressively more complex, complete and mature in knowledge. We build on previous DSR projects and present a design theory by developing its anatomy (Gregor & Jones, 2007), comprising of purpose and scope covered previously with subsequent sections presenting principles of form and function, artefact mutability, testable propositions, justificatory knowledge, principles of implementation and expository instantiation. This explicit attention to theory goes beyond solution development and provides contribution to knowledge for defined health organizational problems (Miah et al., 2019b).

**Justificatory Knowledge**

Justificatory knowledge is *"the underlying knowledge or theory from the natural or social or design sciences that gives a basis and explanation for the design (kernel theories)"* (Piirainen & Briggs, 2011). While we use design thinking to prototype a solution using DSRM the artefact design was grounded in the theories of: systems thinking, and a theory of information need for information retrieval that connects information to knowledge (Cole, 2011).

Systems Thinking is "*a set of synergistic analytic skills used to improve the capability of identifying and understanding systems, predicting their behaviours, and devising modifications to them in order to produce desired effects. These skills work together as a system.*" (Arnold & Wade, 2015). Health sciences libraries in organizations use many systems, the use of which, including optimization, can be quite siloed and unaligned to its larger organizational purpose (Miranda, 2019).

Systems Thinking allows us to identify most systems used by libraries, their interconnections and align them to achieve a common purpose to improve information retrieval holistically instead of via multiple fragmented systems. Different systems are treated as components or elements of a larger system, which we optimize by removing complexity in their inter-connections and configuration.

The premise of a theory of information need is rooted in the 1968 paper by Robert S. Taylor who foresaw (Taylor, 2015) the role of information centres as they would take on a new form at organizations, or these information centres would gradually make the transition from reliance on people to reliance on systems to gather knowledge.

Our research designed the artefact around eight information need concepts grouped by Cole C. into three categories of Information Behaviours, Context and Human Condition (Cole, 2011). This theory of information need for information retrieval reveals systems designed with a computer-science approach are inadequate, and that a system design should use a user-oriented knowledge formulation/acquisition approach.

The two kernel theories described above are connected by the human condition category which requires a holistic approach to understanding the user environment to design user-oriented systems.

**Construct**

From a design theory perspective, constructs represent the entities of interest in a theory. These can be abstract terms which are theoretical or phenomena that are physical or in our case virtual in nature (Gregor & Jones, 2007).

A literature review of systems in use by health sciences libraries for information retrieval, analysis of interviews, and observations of systems found in health organizations, led us to identify two categories of systems – resource providers and resource facilitators based on their nature.

Resource providers are online evidence platforms where the content or full text is hosted for clinicians to retrieve. Resource facilitators are systems clinicians use to journey to full text or content and present an opportunity to improve clinician' research in practice.

Depending on the cloud services procured by an organization, clinicians can use resource facilitators such as WSD (for pain points time, awareness and integration) (Hanneke & O'Brien, 2016), Global Knowledgebase (for pain points resource scope) (Wilson, 2016), Link Resolver (solution for pain point paywall and access) (Chisare et al., 2017), Federated Access (for pain points paywall and access) (Hoy, 2019), CMS (for integration and resource platforms) (De Silva & Burstein, 2014) and Document Delivery or inter-library loan (for pain points paywall, access) (Brian, 2019). The configuration varies significantly from one organization to another. We scoped these systems to remove barriers to information retrieval. A simplified version of the context aware architecture in      Figure 2 illustrates the various constructs as part of the resource facilitators group, and how they are unified using a single interface via an API layer for integration.

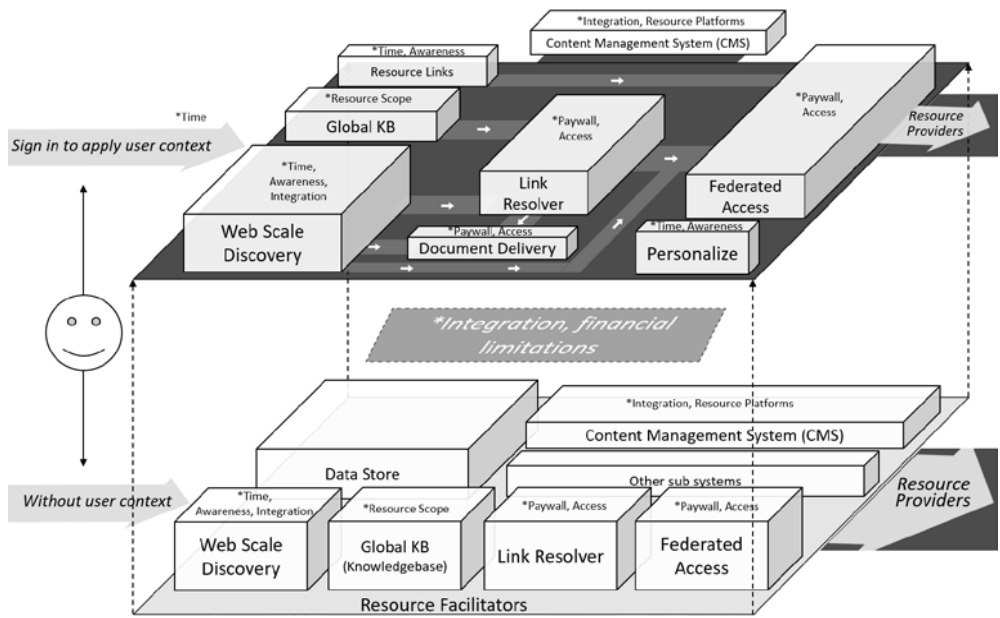

*Figure 2: Context aware unified architecture with resource facilitators*

# RESULTS AND DISCUSSION

## Principles of form, function, and implementation

Principles of form and function in design theory relate to an abstract architecture or "blueprint" that describes an information systems artefact. The context aware unified architecture provides the necessary building blocks to develop an instantiation solution, however, such an architecture is complex to implement and requires underlying principles for general application (Miah et al., 2018). Figure 3 builds on the architecture in Figure 2 by presenting a form and function of the solution and illustrates the principles of implementation.

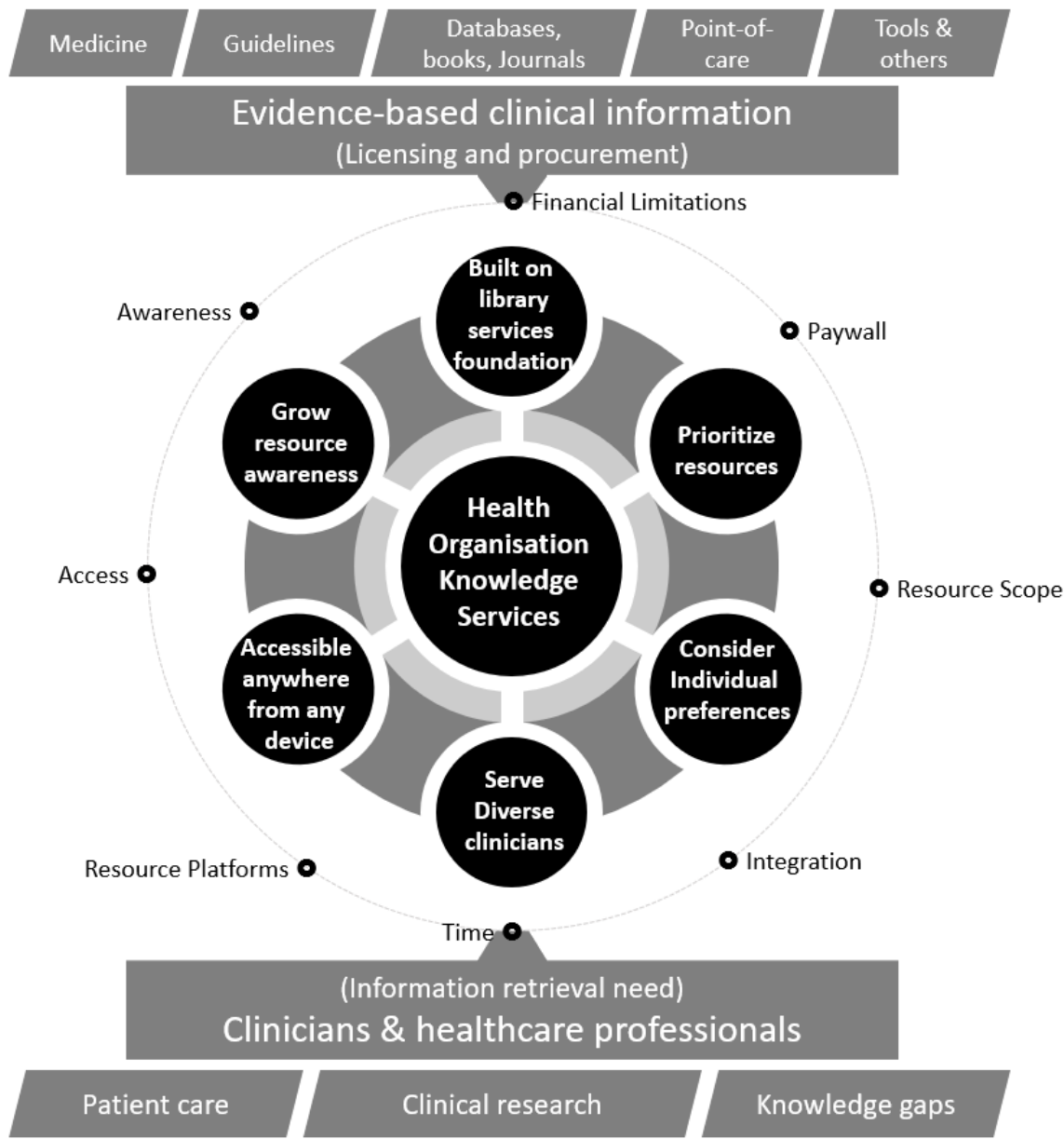

*Figure 3: Design Theory artefact*

Based on DSR fundamentals the bottom section of the illustration in Figure 3 depicts clinicians and healthcare professionals retrieving information for patient care, clinical research and fill knowledge gaps. The upper bar of the illustration depicts the various online resources (e.g. medicine, guidelines, databases, books, journals, point-of-care, tools and other resources), available as evidence-based clinical information for consumption by clinicians. Myriad online resources exist for clinicians to access, or be aware of in what is already a time-poor industry (Miranda & Miah, 2019). Yet knowledge from these resources depending on clinician groups (e.g. nurse, dentist, general practitioner etc.) is essential to evidence-based practice. A barrier exists between a clinician's information retreival need and available evidence-based clinical information, represented by the outer circle orbited by eight pain points. The central circle offers implementation principles for health organizations to design a solution that will both solve the problem and remove the barriers.

All artefacts in this research stream were designed and evaluated using the following implementation principles and the final instantiation in combined form is demonstrated. The design theory illustrates a degree of alignment and proximity between the pain points and principles associated to solving specific pain points.

**Principle 1:** Built on library services foundation. Gardois et. al. (Gardois et al., 2011) and Hall (Hall, 2008) showed that health sciences librarians retrieved information better as they understood their resources. Kamada et. al. (Kamada et al., 2021) reported the close proximity of a physically embedded librarian benefited users. Health sciences librarians play key roles in configuring systems to optimize the solution for their users. An organization can do little to remove barriers on resource provider platforms because these services are designed to serve a wider audience. Resource facilitators such as WSD, Federate Access and CMS can be configured towards knowledge formulation at a health organization level, making such a solution feasible for health sciences librarians to optimize.

Taking stock of all the systems available at a health sciences library in Project B provided the building blocks for a feasible holistic solution (Miranda, 2019). Using these building blocks rooted in library services resonated with library managers to remove barriers while identifying gaps in their services to prioritize future procurement (Miranda & Miah, 2021). Project A used a custom web application as a front-end to prototype contextual information retrieval which required significant development effort (Miranda & Miah, 2019). Using a CMS as front-end for API integration in Project C required significantly less development effort with the added capability for libraries to configure the font-end and better support clinicians' information need.

**Principle 2:** Serve diverse clinicians. Chiu et. al. (Chiu et al., 2012) found that different specialists need to retrieve different types of information and Ragon found that tradition-based service models of health sciences libraries are insufficient to meet the needs of biomedical researchers (Ragon, 2019). An information seeking study by Nwezeh et al. (Nwezeh et al., 2011) indicated that information delivery must be prioritized for clinicians based on their specialized skills information need priorities which may differ by clinician groups and individual clinicians. With hundreds of resources, it is time consuming to use different platforms to research clinical information. Minimizing research time irrespective of clinician's education and skill levels will alleviate delays and ensure that diverse clinician groups are serviced.

Project A identified that an organizational context is applied to services as it aggregates and indexes online evidence containing content across different sectors beyond medical. While this organizational context improved information relevance, an added layer of group context made the results even more relevant to clinicians area of specialization (Miranda & Miah, 2019). Project C built on these findings and demonstrated that health sciences libraries could create different group contexts. This allowed diverse clinicians to borrow or apply their preferred group context as they retrieved information and adjust the borrowed context to suit their individual needs (Miranda & Miah, 2021).

**Principle 3:** Consider individual preferences. Health sciences libraries in organizations can be a one-stop-shop for clinician's information needs, but if a service does not offer a personalized experience, users avoid it and find alternative means to retrieve information. Saparova et. al. (Saparova et al., 2014) piloted a federated search system that factored in information needs of 51 clinicians; key feedback indicated that personalization features improved human, technology, and organization fit. Meeting the individual information needs of clinicians ensures a service delivers value; a lack of this capability was significant feedback in the search system pilot.

Project A highlighted that that while group context significantly improved relevance of results it was important to allow clinicians to remove the group context and discover content at the organizational context level. Similarly, individuals may prefer to tailor group context and organizational context (Miranda & Miah, 2019). Library Managers interviewed during Project C agreed with the results but added that the ability to adjust preferences at a granular level applied to clinicians familiar with library services and the range of options without guidance could overwhelm most clinicians (Miranda & Miah, 2021).

**Principle 4:** Accessible anywhere from any device. Licensed online evidence-based clinical information is typically behind a paywall and requires authentication to access, which can challenge clinicians who seek fast information retrieval in high pressure situations. Shanahan (Shanahan, 2013) reported a lack of access to computers, restrictions on websites, and a lack of time to engage because to heavy workloads limited evidence-based information retrieval. Knowledge delivery should be platform independent and clinicians should be able to consume information at any time and place (Kouame, 2014).

Procuring cloud-native services boosts an organization's ability to let their clinicians access content from anywhere on any device. Most resource providers of evidence-based clinical information allow access via online services making procurement of cloud-native resource facilitators a natural fit for a holistic solution (Miranda, 2019). The time and effort required to release improvements is significantly higher when cloud-native services were not maximized in Project A. Time and effort on IT infrastructure was eliminated in Project C which was optimized to use cloud-native services.

**Principle 5:** Grow resource awareness. Health organizations may subscribe to hundreds of resources, but clinicians limit their use of resources to those with which they are familiar, potentially missing information in other databases, journals, and tools. Davies (Davies, 2011b) demonstrated UK doctor' awareness and use of specific evidence-based medicine resources was limited. With thousands of listed resources, noise can be generated, and knowledge of the relevant resources for individual clinicians and respective groups should be considered when designing systems for information retrieval as recommended by Davies (Davies, 2011a) and Olalekan (Olalekan Moses, 2016).

During Project B we found that an A-to-Z list to browser resources is generally available to clinicians however these are generally exhaustive often used by clinicians to find links to resources, they are familiar with compared to grow resource awareness (Miranda, 2019). In addition to creating group contexts in Project C based on clinician specializations the unified context aware architecture enables health sciences libraries to highlight new or existing resources periodically both at the institutional context and group context levels (Miranda & Miah, 2021). This enables knowledge acquisition or formulation across the solution holistically.

**Principle 6:** Prioritize resources based on clinician needs. Various healthcare professionals have diverse information access needs and only the most relevant resources should be presented depending on the clinicians needs as recommended in research conducted by Chiu et. al. (Chiu et al., 2012) and Davies (Davies, 2011a) who found clinicians information needs vary. Mu et. al. (Mu et al., 2011) evaluated a novel facet view interface based on a topic clustering algorithm and found this improved the user experience compared to a list based interface.

Topic clustering inspired the bento box in Project B which presented a range of information feeds based on group context (Miranda, 2019). Project C introduced context settings at a clinician level allowing individuals to adjust the information feeds they would receive (Miranda & Miah, 2021). Such a knowledge acquisition or formulation solution allows health sciences libraries to present a topic cluster relative to organizational context, another topic cluster relative to group context and any number of topic clusters based on individual preferences to prioritize resources.

**Artifact Mutability**

Artifact mutability as part of a design theory's anatomy refers to *"the changes in state of the artifact anticipated in the theory, that is, what degree of artifact change is encompassed by the theory"* (Gregor & Jones, 2007). The instantiations of models and architectures in this research stream use specific technologies such as WSD, CMS, Global Knowledgebase, Document Delivery as constructs, and the form and function of these technologies evolves over time.

Rather than design principles around specific technological services which might necessitate a computer science approach our research, we refer to the foundational knowledge of user-oriented theory of information need. Irrespective of how technologies might evolve, the design theory and underlying artefacts allow health sciences librarians to adopt cloud-based services to fit their organizational and clinician needs. By using a unified front-end user interface and integration of various services via API, health organizations can add, remove, and change underlying services of search, content management, and knowledgebase document delivery with little impact to the user journey and experience.

Our design theory advocates an iterative and incremental approach to implementation allowing health organizations to adopt resource facilitators at a sustainable pace, compared to a radical change in systems which is also feasible, should a health sciences library prefer this approach.

**Expository instantiation and testable propositions**

An instantiation of a design theory which satisfies testable propositions demonstrate theoretical and practical contribution of the artefact (Piirainen & Briggs, 2011).

We iteratively developed a series of architecture designs, developed instantiations, and demonstrated them to an academic panel with expertise in DSR and information systems until a version satisfied all design principles.

Figure 2 illustrates how the various sub-systems are unified in the one interface using a CMS. The instantiation in Figure 4 include a WSD search box, Federated Access sign in form, and resource links direct to providers including Global Knowledgebase link outs to browse journals, databases, and eBooks. A link under the search field also shows document delivery requests to use when full text is not available.

Upon sign-in clinicians are directed to a personalized section of the CMS where they are recognized as a logged in user. Here a clinician is presented with resource options based on a borrowed group template, or individual preferences as shown in Figure *4*. The prototype selects the general practitioner group context by default for clinicians who have logged in for the first time, however future implementations can be designed with a wizard to walk users through context selection.

Figure *4* integrates the groups context evaluated in Project A but is integrated into a CMS compared to a custom application to ensure feasible adoption by health sciences libraries using cloud services. Pre-defined group templates are General Practitioner, Nurse, Physiotherapist, Medical Researcher, and Psychologist; clinicians choose the group to which they most identify.

This is followed by a My Quick Links block to demonstrate the personalization capabilities of the CMS, where a clinician can pick online evidence-based resources they prefer to work on directly. Some examples include Cochrane Library, PubMed, Lippincott Advisor, CINAHL Complete and Paediatric Care Online.

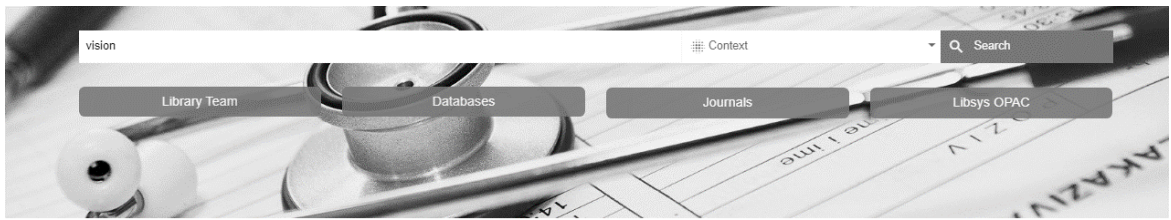

*Figure 4 Prioritized resources view based on borrowed group or individual preferences.*

The layout of the page continues to take on a bento box approach (*sensu* Lown et. al. (Lown et al., 2013)) to determine search result layouts. Here a clinician can choose to view specific highlights from the full data set or limit by source type (e.g., journals, books) and even incorporate publication-level titles in one single view; Title configuration is limited only by the clinician's information needs. Other topic clusters included resources from PubMed and Encyclopedia Britannica referred to as Research Starters. These topic clusters were based on group context with scope to apply user context.

Figure *5* demonstrates the context settings down the page for granular individual preferences. These can be pre-populated depending on the underlying systems, such as CMS taxonomy or facets from WSD services which we adopted.

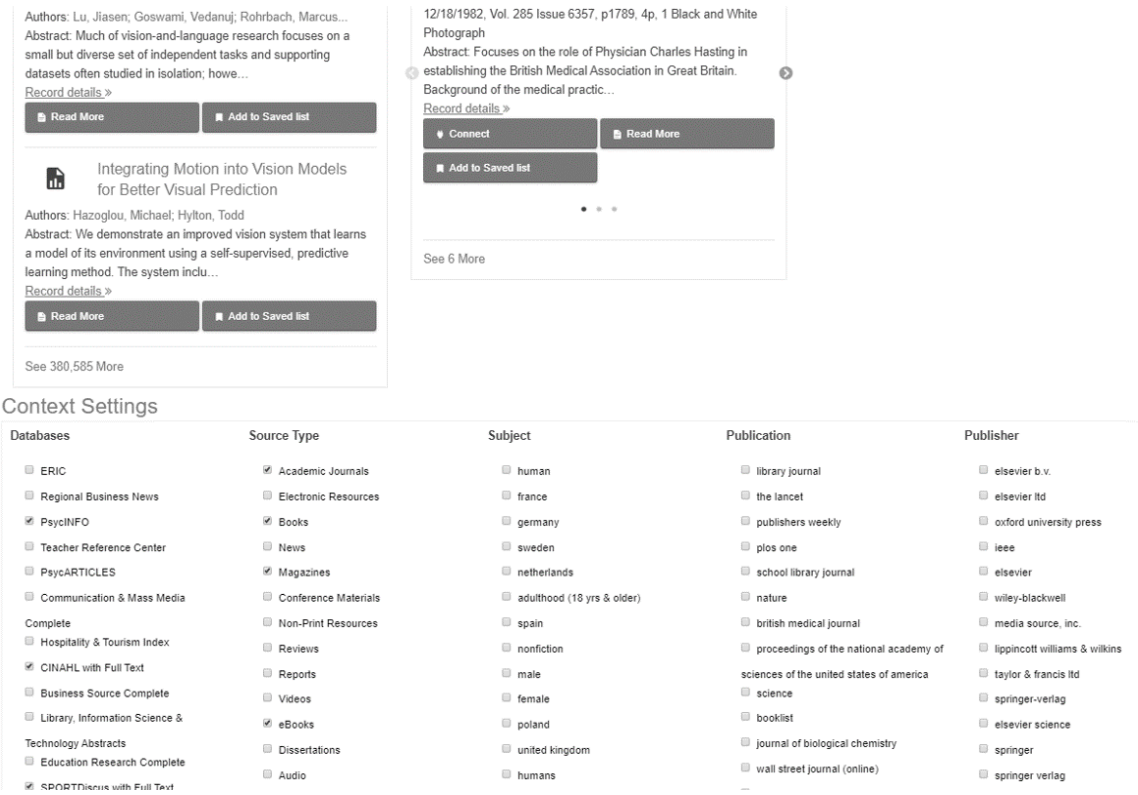

*Figure 5 Context settings to adjust preferences.*

Context settings were designed to appear on the same screen as context selection and bento box results to compare the impact of setting changes to facilitate evaluations. It is up to health sciences libraries to determine where these settings should be presented to users. Instantiation of the settings context is derived from filters from the WSD service to expedite prototype development. Figure *5* shows filters such as Databases, Source Type, Subject, Publication, and Publisher, which are generated automatically based on the facet filters configured in the selected WSD administration tool. Health sciences libraries can enable more or reduce the number of filters based on what is appropriate for clinicians at their respective organizations.

When a user signs into the CMS the context layer queries the user preferences data store to determine if any personalization was applied by the user. If data exists, the selections are applied to the context settings so a user can recommence where they left off configuring their personalization and adjust incrementally. Alternatively, the context layer will retrieve data from the group template store to apply a selected or preferred context.

Figure 6 demonstrates how searching is done in context with facets or filters pre-applied to the results.

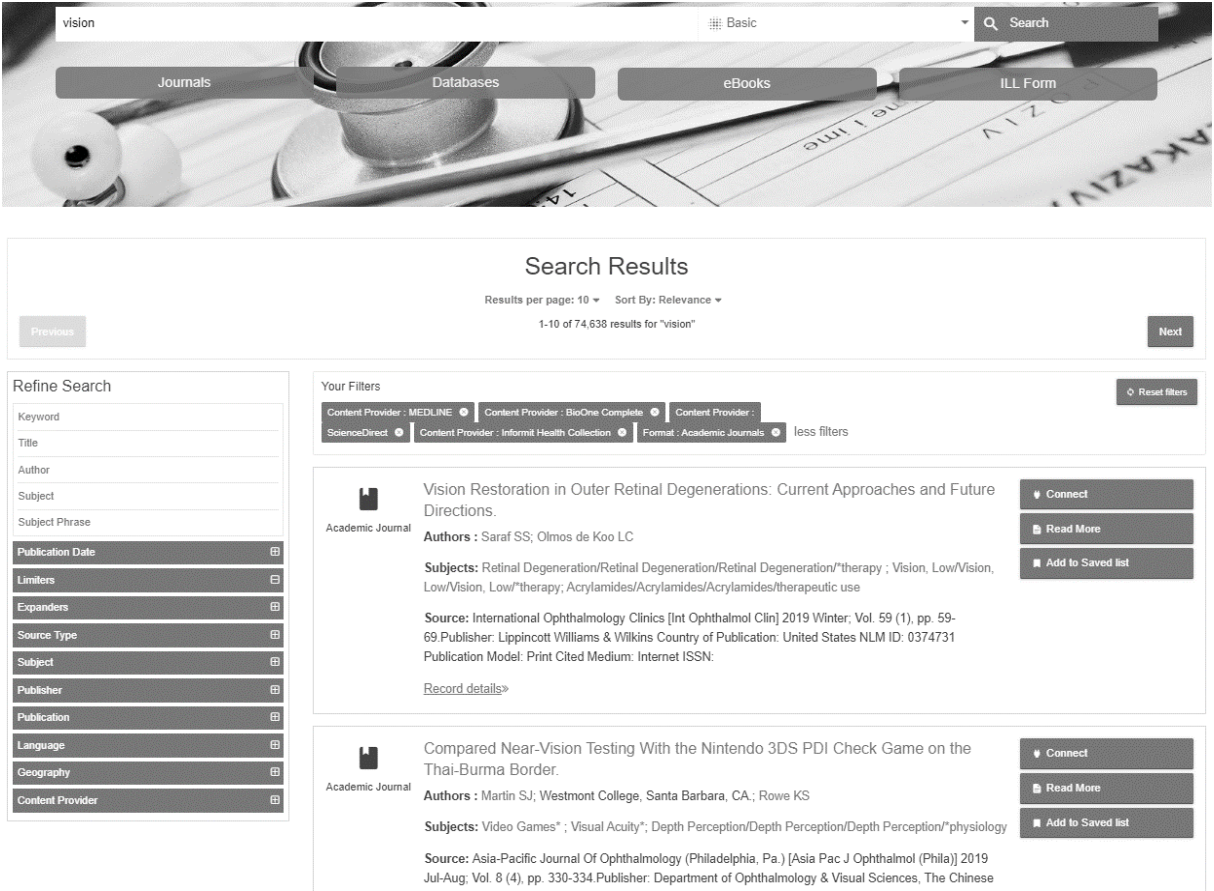

*Figure 6 Search results based on clinician's context.*

Users can enter keywords into the search box at the top of the page. Clicking the search button redirects them to the results page. Search results are presented in a two-column layout with the left column about 30% of the total screen. The layout is responsive and will change depending on the device for maximum usability.

Noteworthy observations in Figure *6* include a section above the results which shows filters such as content providers or databases MEDLINE, BioOne Complete, Science Direct and Format Academic Journals. Users can evaluate the relevance of results to their context by removing filters, achieved by clicking the 'remove' icon next to a filter name. Users can also add filters by expanding the filter types and selecting a filter on the left side of the page.

## CONCLUSION

We conclude that barriers to clinical information retrieval at health organizations can be removed by implementing resource facilitators using the design theory in this paper. Cloud services render a solution significantly feasible and accessible compared to on-premises infrastructure. Health sciences libraries play an important role in designing such a holistic solution to address pain points relative to their organizations.

Artefacts in this research stream were designed and developed after conducting a review of literature wherein problems experienced by researching clinicians were identified, rather than by contacting clinicians directly and soliciting problems. Artefacts were evaluated as useful by an academic panel, clinicians as end users, a confirmatory focus group, and three health sciences library managers in context to their health organizations. We communicated results in our research stream in scholarly publications and conferences, ensuring iterative incremental evaluation from specific instantiations to generalized knowledge by the type "Theory for Design and Action", i.e. theory of how to do things, based on Gregor's types of theories (Gregor, 2006).

Hevner, A. R., March, S. T., Park, J., & Ram, S. (2004). DESIGN SCIENCE IN INFORMATION SYSTEMS RESEARCH [Article]. *MIS Quarterly*, *28*(1), 75-105.

Hoy, M. B. (2019). An Introduction to RA21: Taking Authentication Beyond IP Addresses. *Medical Reference Services Quarterly*, *38*(1), 81-86. https://doi.org/10.1080/02763869.2019.1554370

Jafari Sadeghi, V., Nkongolo-Bakenda, J.-M., Anderson, R. B., & Dana, L.-P. (2019). An institution-based view of international entrepreneurship: A comparison of context-based and universal determinants in developing and economically advanced countries. *International Business Review*, *28*(6), 101588. https://doi.org/https://doi.org/10.1016/j.ibusrev.2019.101588

Johannesson, P., & Perjons, E. (2014). *An introduction to design science* [Book]. Cham : Springer, 2014.

Kamada, H., Martin, J. R., Slack, M. K., & Kramer, S. S. (2021). Understanding the information-seeking behavior of pharmacy college faculty, staff, and students: implications for improving embedded librarian services. *Journal of the Medical Library Association : JMLA*, *109*(2), 286-294. https://doi.org/10.5195/jmla.2021.950

Kasser, J., & Zhao, Y. (2016, 12-16 June 2016). Simplifying solving complex problems. 2016 11th System of Systems Engineering Conference (SoSE),

Kouame, G. (2014). HEALWA: A Unique Information Resource for Washington State Health Professionals: A 5-Year Review. *Journal of Hospital Librarianship*, *14*(4), 391-394. https://doi.org/10.1080/15323269.2014.951219

Laera, E., Gutzman, K., Spencer, A., Beyer, C., Bolore, S., Gallagher, J., Pidgeon, S., & Rodriguez, R. (2021). Why are they not accessing it? User barriers to clinical information access [Information Seeking Behavior; Information Access; User Behavior; Library; Librarian; Publishing; Physician; Clinical Researcher; Clinical Faculty; Paywalls; Financial Limitations; Physicians; User Experience]. *Journal of the Medical Library Association : JMLA*, *109*(1), 7. https://doi.org/10.5195/jmla.2021.1051

Lialiou, P., Pavlopoulou, I., & Mantas, J. (2016). Health Professionals' Use of Online Information Retrieval Systems and Online Evidence. MIE,

Lown, C., Sierra, T., & Boyer, J. (2013). How Users Search the Library from a Single Search Box. *2013*, *74*(3), 15. https://doi.org/10.5860/crl-321

Maria, P., & Shah, J. M. (2019). Blockchain in healthcare. *Australasian Journal of Information Systems*, *23*(0). https://doi.org/10.3127/ajis.v23i0.2203

Miah, S., Gammack, J., & McKay, J. (2018). A Meta-Design Theory for Tailorable Decision Support. *Journal of the Association for Information Systems*.

Miah, S. J., Vu, H., & Gammack, J. (2019a). A big-data analytics method for capturing visitor activities and flows: the case of an island country. *Information Technology and Management*, *20*(4), 203-221. https://doi.org/10.1007/s10799-019-00303-2

Miah, S. J., Vu, H. Q., & Gammack, J. G. (2019b). A Location Analytics Method for the Utilisation of Geotagged Photos in Travel Marketing Decision-Making. *Journal of Information & Knowledge Management*, *18*(01), 1950004. https://doi.org/10.1142/S0219649219500047

Miranda, A. (2019, 2/09/2019). *Unified library services: an architecture to improve research evidence through digital medical libraries* AMLEE-2019, National Seminar on Access &

Taylor, R. S. (2015). Question-Negotiation and Information Seeking in Libraries. *2015*, *76*(3), 17. https://doi.org/10.5860/crl.76.3.251

Venable, J., Pries-Heje, J., & Baskerville, R. (2016). FEDS: a framework for evaluation in design science research. *European Journal of Information Systems*, *25*(1), 77-89.

Wilson, K. (2016). The Knowledge Base at the Center of the Universe [Article]. *Library technology reports*, *52*(6), 1-35.
24